\documentclass[12pt]{iopart}

\usepackage{graphicx}
\usepackage{color}
\usepackage{mathrsfs}
\usepackage{bm}
\usepackage{bbm}
\usepackage{amsfonts}
\begin{document}

\title{Optimal control of non-Markovian open quantum systems via feedback}
\author{Zairong Xi$^{1 *}$, Wei Cui$^{1,2}$, and Yu Pan$^{1,2}$}
\address{%
$^1$Key Laboratory of Systems and Control, Institute of Systems
Science, Academy of Mathematics and Systems Science, Chinese Academy
of Sciences, Beijing 100190, People's Republic of China

$^2$Graduate University of Chinese Academy of Sciences, Beijing
100039, People's Republic of China
}%
\ead{zrxi@iss.ac.cn}

\begin{abstract}
The problem of optimal control of non-Markovian open quantum system
via weak measurement is presented.
Based on the non-Markovian master equation, we evaluate exactly the
non-Markovian effect on the dynamics of the system of interest
interacting with a dissipative reservoir. We find that the
non-Markovian reservoir has dual effects on the system: dissipation
and backaction. The dissipation  exhausts the coherence of the
quantum system, whereas the backaction revives it. Moreover, we
design the control Hamiltonian with the control laws attained by the
stochastic optimal control and the corresponding optimal principle.
At last, we considered the exact decoherence dynamics of a qubit in
a dissipative reservoir composed of harmonic oscillators, and
demonstrated the effectiveness of our optimal control strategy.
Simulation results showed that the coherence  will  completely lost
in the absence of control neither in non-Markovian nor Markovian
system. However, the optimal feedback control steers it to a
stationary stochastic process which fluctuates around the target. In
this case the decoherence can be controlled effectively, which
indicates that the engineered artificial reservoirs with optimal
feedback control may be designed to protect the quantum coherence in
quantum information and quantum computation.

\end{abstract}

\submitto{\JPA}
\maketitle

\section{Introduction}

Quantum information science has emerged as one of the most exciting
scientific developments of the past decade. The most difficult
problem in realizing the quantum information technology is that the
quantum system can never be isolated from the surrounding
environment completely. Interactions with the environment
deteriorate the purity of quantum states. This general phenomenon,
known as decoherence \cite{Breuer,weiss}, is a serious obstacle
against the preservation of quantum superpositions over long periods
of time. Decoherence entails nonunitary evolutions, with serious
consequences, like a loss of information and probable leakage toward
the environment.
Thus, on the one hand, the information carried by a quantum system
has to be protected from decoherence. On the other hand, a detailed
knowledge of quantum dynamics and control will help to construct
high precision devices that accurately perform their intended tasks
despite large disturbances from the environment.
%

For the purpose of efficiently processing quantum information,
results and methods coming from the classical control theory have
been systematically introduced to manipulate practical quantum
systems
\cite{Zhang1,Zhang09,ZhangJ:07,Wiseman:93,Rabitz:00,Wiseman:10,Wu:06,ZhangM:06,Cui,Pan,Xi:
08}.
 Early works aim to answer
the questions like controllability and observability
 in finite dimensional closed quantum
systems \cite{Tarn:80,Belavkin:83}. They have many applications like
molecular dynamics in quantum chemistry. As we demonstrated above,
quantum systems are very vulnerable when exposed in a noisy
environment. So, open quantum systems control is becoming a matter
of concern for more and more physicists and cyberneticists.
Recently, quantum control works include: (i) steering a quantum
system from its initial state to a given final state or a set of
final states \cite{Wei:08}. Transferring to specific final states
are very important for applications to quantum computing, quantum
chemistry and atomic physics. (ii) Quantum decoherence control
\cite{Zhang1,Cui}. Suppression of decoherence is essential to
effectively protect quantum purity and quantum coherence. (iii)
Quantum entanglement control \cite{Cui:091,Cui:092}. Reliable
generation and distribution of entanglement in a quantum network is
a central subject in quantum information technology, especially in
quantum communication.

To the quantum control strategies, a number of interesting schemes
have been proposed during the last few years in order to protect the
purity and counter the effects of decoherence. It is just like the
classical control theory, according to the principle of controllers'
design, the open quantum systems control includes open-loop control
and closed-loop control. Quantum open-loop control means to design
the controller without measurement. Early proposed quantum
error-correction codes, error-avoiding codes and Bang-Bang control
 can be classified as open-loop control. Nowadays, by
tuning the system's Hamiltonian, coherent control theory has opened
new perspective on decoherence and entanglement control. It can
decouple a part of the system dynamics from decoherence, on which
noiseless quantum information can be encoded either in Markovian
open quantum systems or in non-Markovian open quantum system
\cite{Maniscalco1,Maniscalco2}. The quantum coherent control plays
an important role in state generation and transfer in quantum
information, quantum chemistry and optical physical, and it is a
focus these days. Feedback is the essential concept of classical
control theory, but
 the study of feedback control has only a toehold in quantum
 control theory  for many fundamental problems needed to be solved.
 The main problem is quantum measurement. Feedback control needs
 measurement to reduce the system's uncertainty, whereas measuring a
 quantum system will inevitably lead to quantum state collapse,
 which increase the system's uncertainty in turn \cite{Qi}. As we
  demonstrated
 above, the theory of quantum measurement is strange in
 that it does not allow the noncommuting observables to be measured
 simultaneously.
The essence of feedback control is expected to attain robustness to
noise or modeling error, and quantum control using continuous
measurement, so-called measurement based quantum feedback control,
was proposed in  early 90's.  Recently, the stochastic control
theory has been exploited to the open quantum system. The theory of
quantum feedback control with continuous
 measurement can be developed simply by replacing each
 ingredient of stochastic control theory by its noncommutative
 counterpart. The system and observations are described
by quantum stochastic differential equations. The field of quantum
stochastic control was pioneered by V. P. Belavkin in the remarkable
 paper \cite{Belavkin:92} in which the quantum counterparts
of nonlinear filtering and LQG control were developed. The advantage
of the quantum stochastic approach is that the details of quantum
probability and measurement are hidden in a quantum filtering
equation. In view of physicists, they also reformulated the
evolution of damped systems with output in the form of an explicitly
stochastic evolution equation, which specifies the quantum
trajectory of the systems. Feedback control of quantum mechanical
systems which take into account the probabilistic nature of quantum
measurement is one of the central problem in the control of such
systems in both physics and control theory. Recently,
superconducting qubits \cite{You:05,Liu:05} have also been studied
as ways to control and interact with naturally formed quantum
two-level systems in superconducting circuits. The two-level systems
naturally occurring in Josephson junctions constitute a major
obstacle for the operation of superconducting phase qubits. Since
these two-level systems can possess remarkably long decoherence
times, References \cite{Ashhab1,Ashhab2,Ashhab3} showed that such
two-level systems can themselves be used as qubits, allowing for a
well controlled initialization, universal sets of quantum gates, and
readout. Thus, a single current-biased Josephson junction can be
considered as a multi-qubit register. It can be coupled to other
junctions to allow the application of quantum gates to an arbitrary
pair of qubits in the system. These results
\cite{Ashhab1,Ashhab2,Ashhab3} indicate an alternative way to
control qubits  coupled to naturally formed quantum two-level
systems, for improved superconducting quantum information
processing.  Indeed, these predictions have been found
experimentally in \cite{Neely:08}. More recently, reference
\cite{Burgarth:09} applies quantum control techniques to control a
large spin chain by only acting on two qubits at one of its ends,
thereby implementing universal quantum computation by a combination
of quantum gates on the latter and swap operations across the chain.
They \cite{Burgarth:09} show that the control sequences can be
computed and implemented efficiently. Moreover, they discuss the
application of these ideas to physical systems such as
superconducting qubits in which full control of long chains is
challenging.

The rest of the paper is organized as follows. In Section II, we
introduce the open quantum system and the non-Markovian quantum
master equation, and present the main difference between the
non-Markovian quantum system and the Markovian one. The optimal
control problem via quantum measurement feedback is precisely
formulated in Section III. In Section IV, we specifically
investigated the stochastic optimal control of Spin-Boson system.
 A useful numerical example is demonstrated in Section V,
and a few concluding remarks are given in Section VI.

%
%
%

\section{Open quantum system and non-Markovian master
equation}

 The theory of open quantum system describes the dynamics
of a system of interest interacting with its surrounding environment
\cite{Breuer}, and the quantum master equation governs the evolution
of the quantum system, which plays an important role in relaxation
and decoherence theory. Markovian approximation is usually used in
this master equation under the assumption that the correlation time
between the systems and environments is infinitely short. However,
in some cases an exactly analytic description of the open quantum
system dynamic is needed. Especially in high-speed communication the
characteristic time scales become comparable with the reservoir
correlation time, and in solid state devices memory effects are
typically non negligible. So it is necessary to extensively study
the non-Markovian case. In this paper, we will consider the optimal
control of non-Markovian open quantum system via feedback, typically
the quantum decoherence optimal control.

\subsection{Open quantum system}

 Consider a quantum system $S$ embedded in a
dissipative environment $B$ and interacting with a time-dependent
external field, i.e., the control field. The total Hamiltonian has
the general form
\begin{equation}
\begin{array}{rcl}
\hat{H}_{tot}&=&\hat{H}_S+\hat{H}_{B}+\hat{H}_{I}\\
&=&\hat{H}_0+\hat{H}_{C}(t)+\hat{H}_{B}+\hat{H}_{I},
\end{array}
 \end{equation}
where $\hat{H}_{0}$ is the system free Hamiltonian, $\hat{H}_{C}(t)$
the Hamiltonian of the control field (either open-loop or
closed-loop), and $\hat{H}_{B}$ the bath and $\hat{H}_{I}$ the
interaction between the free system and the bath. Note that the
interaction between the control field and the free system or the
bath are always negligible. The operator $\hat{H}_{C}(t)$ contains a
time-dependent external field to adjust the quantum evolution of the
system. One of the central goals of the theoretical treatment is
then the analysis of the dynamical behavior of the populations and
coherences, which are given by the elements of the reduced density
matrix, defined as
\begin{equation}
\rho_{S}(t)=\tr_B[\rho_{tot}(t)],
 \end{equation}
where $\rho_{tot}$ is the total density matrix for both the system
and the environment. For simplicity, it is always assumed as the
system of a single atom which configured such that a transition
between only two states is possible.  The Hamiltonian of the
environment is assumed to be composed of harmonic oscillators with
natural frequencies $\omega_i$ and masses $m_i$,
 \begin{equation}
   \hat{H}_{B}=\sum_{i=1}^{N}\left(\frac{p_i^2}{2m_i}+\frac{m_i}{2}x_i^2\omega_i^2\right),
 \end{equation}
 where $(x_1,x_2,\cdots ,x_N,p_1,p_2,\cdots, p_N)$ are the
 coordinates and their conjugate momenta, and the Planck constant
 $\hbar=1$ in atomic units and the initial state of the system is
  $\rho(0)=\rho_0$ (for simplicity we write $\rho$ as $\rho_S$).
  For convenience, we always assume that the evolution starts from
  $t_0=0$.
 The interaction Hamiltonian between the system $S$ and the environment $B$
 is assumed to be bilinear,
\begin{equation}
   \hat{H}_{int}=\alpha\sum_{n}A_{n}\otimes B_{n}.
 \end{equation}
The interaction Hamiltonian in the interaction picture therefore
takes the form
\begin{equation}
\begin{array}{rcl}
\hat{H}_{int}(t)&=&e^{i(H_S+H_B)t}H_{int}e^{-i(H_S+H_B)t}\\
&=&\alpha\sum_nA_n(t)\otimes B_n(t),
\end{array}
 \end{equation}
where \[
\begin{array}{rcl}
A_n(t)&=&e^{iH_St}A_ne^{-iH_St},\\
B_n(t)&=&e^{iH_Bt}B_ne^{-iH_Bt}.
\end{array}\]
The
 effect of the environment on the dynamics of the system can be
seen
 as a interplay between the dissipation and fluctuation phenomena. And it
 is the general environment that makes the quantum system lose
 coherence (decoherence).
 The system-environment interaction leads to
non-unitary reduced system dynamics.

\subsection{Quantum master equations}

 The
analysis of the time evolution of open quantum system plays an
important role in many applications of modern physics. With the
Born-Markovian approximation the dynamics is governed by a master
equation of the relatively simple form
\begin{equation}
\label{Master equation1} \frac{d}{d
t}\rho(t)=-i[\hat{H}_S(t),\rho(t)]+\sum_m \gamma_m\mathcal
{D}[C_m]\rho(t),
\end{equation}
with a time-independent generator in the Lindblad form, her the
superoperator $\mathcal {D}[L]\rho=L\rho
L^{\dag}-\frac{1}{2}L^{\dag}L\rho-\frac{1}{2}\rho L^{\dag}L$.
 This is the most general form for the generator of a quantum
 dynamical semigroup. The Hamiltonian $\hat{H}_S(t)$ describes the coherent
 part of the time evolution. Non-negative quantities $\gamma_m$ play the role of relaxation rates
 for the different decay modes of the open system. The operators $C_m$ are usually
 referred to as Lindblad operators which represent the various
 decay modes, and the corresponding density matrix equation (\ref{Master equation1}) is
 called the Lindblad master equation. The solution of Eq. (\ref{Master
 equation1}) can be written in terms of a linear map $V(t)=\exp(\mathcal
 {L}t)$ that transforms the initial state $\rho(0)$ into the state
 $\rho(t)=V(t)\rho(0)$ at time $t$. The physical interpretation of
 this map $V(t)$ requires that it preserves the trace and the
 positivity of the density matrix $\rho(t)$.

The most important physical assumption which underlies the Eq.
(\ref{Master equation1}) is the validity of the Markovian
approximation of short environmental correlation times
\cite{Cui:091}. With this approximation, the environment acts as a
sink for the system information. Due to the system-reservoir
interaction, the system of interest loses information on its state
into the environment, and this lost information does not play any
further role in the system dynamics. However, if the environment has
a non-trivial structure, then the seemingly lost information can
return to the system at a later time leading to non-Markovian
dynamics with memory. This memory effect is the essence of
non-Markovian dynamics  \cite{Wiseman:08,Piilo:08,Wolf:08}, which is
characterized by pronounced memory effects, finite revival times and
non-exponential relaxation and decoherence. Non-Markovian dynamics
plays an important role in many fields of physics, such as quantum
optics, quantum information, quantum chemistry process, especially
in solid state physics. As a consequence the theoretical treatment
of non-Markovian quantum dynamics is extremely demanding. However,
in order to take into account quantum memory effect, an
integro-differential equation is needed which has complex
mathematical structure, thus prevent generally to solve the dynamics
of the system of interest. An appropriate scheme is the
time-covolutionless (TCL) projection operator technique which leads
to a time-local first order differential equation for the density
matrix.

 The general structure of the  TCL master equation is given
by
\begin{equation}
\label{Master equation2} \frac{d}{d
t}\rho(t)=-i[\hat{H}_S(t),\rho(t)]+\sum_m\Delta_m(t)\mathcal
{D}[C_m(t)]\rho(t).
\end{equation}
 The first
term describes the unitary part of the evolution. The latter
involves a summation over the various decay channels labeled by $m$
with corresponding time-dependent decay rates $\Delta_m(t)$ and
arbitrary time-dependent system operators $C_m(t)$.

In the simplest case the rates $\Delta_m$ as well as the Hamiltonian
$\hat{H}_S$ and the operators $C_m$ are assumed to be
time-independent, that is, it is the Markovian case. Note that, for
arbitrary time-dependent operators $\hat{H}_S(t)$ and $C_m(t)$, and
for $\Delta_m(t)\geq0$ the generator of the master equation
(\ref{Master equation2}) is still in Lindblad form for each fixed
time $t$, which may be considered as time-dependent quantum
Markovian process. However, if one or several of the $\Delta_m(t)$
become temporarily negative which expresses the presence of strong
memory effects in the reduced system dynamics, the process is then
said to be non-Markovian.

\subsection{Quantum feedback control}
Quantum feedback control was formally initiated by Belavkin's work
\cite{Belavkin:83}, \cite{Belavkin:92} and so on. The main problem
of optimal quantum feedback control is separated into quantum
filtering which provides optimal estimates of the stochastic quantum
operators and then an optimal control problem based on the output of
the quantum measurement. As we demonstrated in the introduction,
 the quantum system is not directly observable.
 Since the very beginning of the quantum mechanics, the measurement
process has been a most fundamental issue. The main characteristic
feature of the quantum measurement is that the measurement changes
the dynamical evolution. This is the main difference of the quantum
measurement compared to its classical analogue. When one observes
 incompatible
 quantum events (self-adjoint orthoprojectors with $P^2=P=P^*$ ($*$ denotes the Hermitian adjoint) acting
 in some Hilbert space $\mathcal {H}$), the state of the system
 needs to be updated to account for the change to the system or
 back-action. This state change was traditionally described by the
 normalized projection postulate
 $$\rho\rightarrow\rho_i=\frac{P_i\rho P_i}{\tr[\rho
 P_i]},$$
 which also ensures instantaneous repeatability of the observed
 event corresponding to the projection $P_i$.

 We now couple the system to a control field. If
we assume no scattering between the measurement and control fields
and assume a weak coupling such that information is not lost into
the control field, then this effectively replaces the Hamiltonian
$\hat{H}_S(t)$ of the system with a controlled Hamiltonian
$\hat{H}_0(t)+\hat{H}(u(t))$ for admissible real valued control
functions $u(t)\in \mathbb{R}$. The term feedback refers to control
law $u(t)$, which is also a function of the state $\rho$. This
control Hamiltonian generates the controlled unitaries $U_t(u(t))$
given the controlled flow $U_t^*(u_{t_0}^T)\rho U_t(u_{t_o}^T)$,
where $u_{t_0}^T\equiv\{u(t)|t_0\leq t<T\}$ is the control process
over the interval $[0,T)$. In classical control, we can allow
complete observability of the controllable system, so that feedback
controls are determined by the system variables. However, in quantum
system the measurement are weak measurement (for non-observability
of the system operators). So, the feedback controls should be given
by a function of the stochastic output process $W_{F_i}(t)$, and
measurement trajectory is fed into the controls
$W_{F_i}(t)\rightarrow u_t(W_{F_i}(t))$.

Considering both controls on the systems and simultaneous quantum
weak measurements over multiple observable $\{F_i\}$, we have
following stochastic master equation and measurement outputs for the
quantum feedback control system:
\begin{eqnarray}
d\rho&=&-i[H_0,\rho]dt+\sum_{k=1}^{r}u_k(t)(-i[H_k,\rho])dt+\sum_m\Delta_m(t)\mathcal
{D}[C_m]\rho dt\nonumber\\
&& +\sum_iM_i\mathcal {D}[F_i]\rho dt+\sum_i\sqrt{M_i\eta_i}\mathcal
{H}[F_i]\rho dW_{F_i}\label{SME1}\\
 dW_{F_i}&=&dY_i-\sqrt{\eta_iM_i} \tr\{F_i\rho\}dt
\end{eqnarray}
where $H_k,~~k=1,2,\cdots,r$ are the control Hamiltonians adjusted
by the time-dependent control laws $u_k(t)\in \mathbb{R}$ and the
superoperator $\mathcal {H}[A]\rho=A\rho+\rho A-[\tr(A\rho+\rho
A)]\rho$. Note that, the last two terms in Eq. (\ref{SME1}) come
from the measurement-induced disturbance (we will explain the
degenerate in the next section from the quantum measurement view of
point), where $M_i$ are the effective interaction strengths and
$\eta_i$ are the detection efficiency coefficients. $Y_i$ are the
observation process. In fact, the innovation process $W_{F_i}$ is a
Wiener process, which satisfy following rules. The expectation of
any integral over $dW_{F_i}$ or $dW_{F_i}^*$ vanishes. The
differentials $dW_{F_i}$ , $dW_{F_i}^*$ commute with any adapted
process. Finally, the quantum It\^{o} rules are
$$dW_{F_i}dW_{F_j}^*=\delta_{F_i,F_j}dt,~~dW_{F_i}^2=(dW_{F_i}^*)^2=dW_{F_i}^*dW_{F_i}=0.$$

\section{Stochastic optimal control of Spin-Boson system}

We have described the quantum feedback controls and have obtained
controlled master equations, the remaining question is: how do we
choose the control strategy $u(t)$ to achieve a particular control
goal? For a fixed function $u(t)$ the expression (\ref{SME1})
becomes a stochastic differential equation and its solution
$\rho^*(t)$ is called a trajectory of the quantum system relevant to
the quantum  control $u(t)$. In general, any trajectory $\rho^*(t)$
should fulfill some initial conditions. In this section we approach
this problem using optimal control theory, where the control goal is
expressed as the desire to minimize a certain cost functional
$J(t,u,\rho)$.

The optimality of control is judged  by the expected cost associated
to the admissible control process $u_{t_0}^T$ for the finite
duration $T$ of the experiment. Admissible control strategies are
defined as those $u_{t_0}^T$ for which the operator valued cost
integral
\begin{eqnarray}
J[u(\cdot)]=\int_{t_0}^TL(\tau,u,\rho)d\tau+\theta \cdot
g(\rho_T,\rho^*),
\end{eqnarray}
and the solutions of Eq. (\ref{SME1}) exist.
 Here $L$ is a function of time $t$, the current control law $u$, and
the quantum state $\rho$. The function $g$, known as a target or
bequest function in control theory, is a function of the state
parameters at termination. We assume that both are continuous in
their arguments. $\theta$ is the constant parameters to adjust the
penalty functional between the two functions. The cost $J$ will vary
from one experimental trial to another, and must be thought of as a
random variable depending on the measurement output.

Here we consider a situation in which a single qubit interacts with
a heat bath of bosons.  We will make use of the two-level
non-Markovian master equation.  The kinetic equation of two-level
strong coupling non-Markovian quantum system is the particular form
of the Eq. (\ref{Master equation2}) and (\ref{SME1}), which
performed a rotating wave approximation after tracing over the
environment without neglecting the counter-rotating terms. By
tracing out the bath degrees of freedom, we find for $\rho$ the
feedback controlled two-level quantum system non-Markovian evolution
equation
%

\begin{eqnarray}
\label{SME2}
d\rho&=&-\frac{i}{2}\omega_0[\sigma_z,\rho]dt-\frac{i}{2}u_x(t)[\sigma_x,\rho]dt-\frac{i}{2}u_y(t)[\sigma_y,\rho]dt\nonumber\\
&+&\Gamma_1(t)\mathcal {D}[\sigma^{-}]\rho dt+\Gamma_2(t)\mathcal
{D}[\sigma^{+}]\rho dt+ M\mathcal {D}\left [-\frac{\sigma_z}{2} \right ]\rho dt\nonumber\\
&+&\sqrt{\eta M}\mathcal {H}\left [-\frac{\sigma_z}{2} \right ]\rho
dW\equiv f(t)
\end{eqnarray}
The free Hamiltonian of the one-qubit quantum system can be written
as
$H_0=\frac{\omega_0}{2}(|0\rangle\langle0|-|1\rangle\langle1|)=\frac{\omega_0}{2}\sigma_z$,
where $|0\rangle$ and $|1\rangle$ are Dirac notations of its two
eigenstates and $\omega_0$ is the Rabi frequency.
$\sigma^+=\frac{1}{2}(\sigma_x+i\sigma_y),~~\sigma^-=\frac{1}{2}(\sigma_x-i\sigma_y)$,
with $\sigma_x$, $\sigma_y$, $\sigma_z$ the Pauli matrices. Here,
the controlled part has two channels
$H_{C}(t)=\frac{1}{2}u_x(t)\sigma_x+\frac{1}{2}u_y(t)\sigma_y$.
The effective interaction strengths between the measurement and the
system is $M\geq0$.

Suppose $\rho=\frac{1}{2}\left(\begin{array}{cc}
1+z&x-iy\\
x+iy&1-z
\end{array}\right)$, thus corresponding to (\ref{SME2}), we have the
following evolution equations,
\begin{eqnarray}\label{SME3}
dx&=&\left(-\Delta(t)x-\frac{M}{2}x-\omega_0y+u_y(t)z\right)dt+\sqrt{M\eta}xzdW\nonumber\\
dy&=&\left(\omega_0x-\Delta(t)y-\frac{M}{2}y-u_x(t)z\right)dt+\sqrt{M\eta}yzdW\nonumber\\
dz&=&2\left(-\frac{u_y(t)}{2}x+\frac{u_x(t)}{2}y-\Delta(t)z-\gamma(t)\right)dt\nonumber\\&&+\sqrt{M\eta}(-1+z^2)dW.
\end{eqnarray}
Thus the coherence factor $\Lambda(t)=\sqrt{x^2+y^2}/2$, and the
population $\rho_{00}(t)=(1+z)/2,~\rho_{11}(t)=(1-z)/2$.

 Here, the objective is to compute an appropriate feedback control
functions $u_x(t),~~u_y(t)$ steering the system from the initial
state $\rho_0$ into a target state $\rho_T$ at final time $T$. The
relative simple cost functional may be written as
\begin{eqnarray}
J[u(t)]=\frac{\theta}{2}\|\rho(T)-\rho_T\|^2+\frac{1}{2}\int_{t_0}^T(u_x^2(t)+u_y^2(t))dt,
\end{eqnarray}
where $\|\cdot\|$ is the Frobenius norm: $\|A\|^2=\tr
A^{\dag}A=\sum_{ij}|A_{ij}|^2$. Here, the first term represents the
deviation between the state of the system at final time $\rho(T)$
and the target state $\rho_T$, the second integral term penalizes
the control field with $\theta>0$, the weighting factor used to
achieve a balance between the tracking precision and the control
constraints. Minimizing the first term is equivalent to maximizing
the state transfer fidelity $F=\tr\{\rho(T)\rho_T\}$. Our overall
task is to find control $u_x(t),~~u_y(t)$ that minimizes $J[u(t)]$
and satisfies both dynamic constraint and boundary condition. The
optimal solution of this problem will be obtained using the optimal
principle. The corresponding Hamiltonian function may be present in
the form
\begin{eqnarray}\label{ha}
\mathcal
{H}(\rho,u,\lambda)=\mathbb{E}\left\{\frac{1}{2\theta}(u_x^2+u_y^2)+\lambda^T(t)f(t)\right\},
\end{eqnarray}
where the adjoint state variable
$\lambda(t)=[\lambda_1(t),\lambda_2(t),\lambda_3(t)]^T$ is the
Lagrange multipliers introduced to implement the constraint. The
optimal solution can be solved by the following differential
equation with two-sided boundary values:
\begin{eqnarray}
\dot{\rho}&=&\frac{\partial\mathcal
{H}}{\partial\lambda},~~~~~~\rho(t_0)=\rho_0;\label{o1}\\
\dot{\lambda}&=&-\frac{\partial\mathcal
{H}}{\partial\rho},~~~~~~\lambda(T)=\rho(T)-\rho_T.\label{o2}
\end{eqnarray}
Together with
\begin{eqnarray}
\left.\frac{\partial \mathcal{H}}{\partial
u}\right|_*&=&\frac{\partial
\mathcal{H}(\rho^*(t),u^*(t),\lambda(t),t)}{\partial
 u}=0;\\
\left.\frac{\partial^2\mathcal{H}}{\partial
u^2}\right|_*&=&\frac{\partial^2\mathcal{H}(\rho^*(t),u^*(t),\lambda(t),t)}{\partial
 u^2}\leq0.
\end{eqnarray}

Take this into Eq.(\ref{SME3}), we get the optimal control for the
 Spin-Boson system with Hamiltonian function Eq.(\ref{ha})
\begin{eqnarray}
 u_x(t)&=&\frac{\theta}{2}\left\{\lambda_2(t)x_3(t)-\lambda_3(t)x_2(t)\right\},\label{o3}\\
 u_y(t)&=&\frac{\theta}{2}\left\{\lambda_3(t)x_1(t)-\lambda_1(t)x_3(t)\right\}.\label{o4}
\end{eqnarray}

 The influence of environmental disturbances include both
 relaxation and dephasing effects, which are formulated as $\Gamma_1(t)\mathcal
 {D}[\sigma^{-}]\rho$ and $\Gamma_2(t)\mathcal
{D}[\sigma^{+}]\rho$.
 The time dependent
coefficients $\Gamma_1(t)=\Delta(t)+\gamma(t)$ and
$\Gamma_2(t)=\Delta(t)-\gamma(t)$,
 with $\Delta(t)$ and $\gamma(t)$ being the
diffusive term and damping term respectively, which can be written
up to the second order in the system-reservoir coupling constant, as
follows,
\begin{equation}
\nonumber \label{Delta}
 \Delta(t)=\int_0^td\tau k(\tau)\cos(\omega_0\tau),
 \end{equation}
\begin{equation}
\label{gamma}
 \gamma(t)=\int_0^td\tau \mu(\tau)\sin(\omega_0\tau),
 \end{equation}
with
\begin{equation}
\nonumber
 k(\tau)=2\int_0^{\infty}d\omega
J(\omega)\coth[\hbar\omega/2k_BT]\cos(\omega \tau), \label{k}
 \end{equation}
\begin{equation}
\mu(\tau)=2\int_0^{\infty}d\omega J(\omega)\sin(\omega \tau),
\label{mu}
 \end{equation}
being the noise and the dissipation kernels, respectively. The
properties of the Eq.~(\ref{SME2}) strongly depend on the behavior
of the dissipation and the noise kernel which, in turn, is
determined by the spectral density $J(\omega)$. In order to obtain
true irreversible dynamics one introduces a continuous distribution
of bath modes and replaces the spectral density by a smooth function
of the frequency $\omega$ of the bath modes. In particular we will
consider  the Ohmic spectral density with a Lorentz-Drude cutoff
function,
\begin{equation}
J(\omega)=\frac{2\gamma_0}{\pi}\omega\frac{\omega_c^2}{\omega_c^2+\omega^2},
 \end{equation}
 where $\gamma_0$ is the frequency-independent damping constant and
 usually assumed to be $1$.  $\omega_c$ is the high-frequency cutoff.
The analytic expression for the coefficients $\gamma(t)$and
$\Delta(t)$ are given in \cite{Cui,Cui:092}.

 \begin{figure}
 \centerline{\scalebox{1}[0.9]{\includegraphics{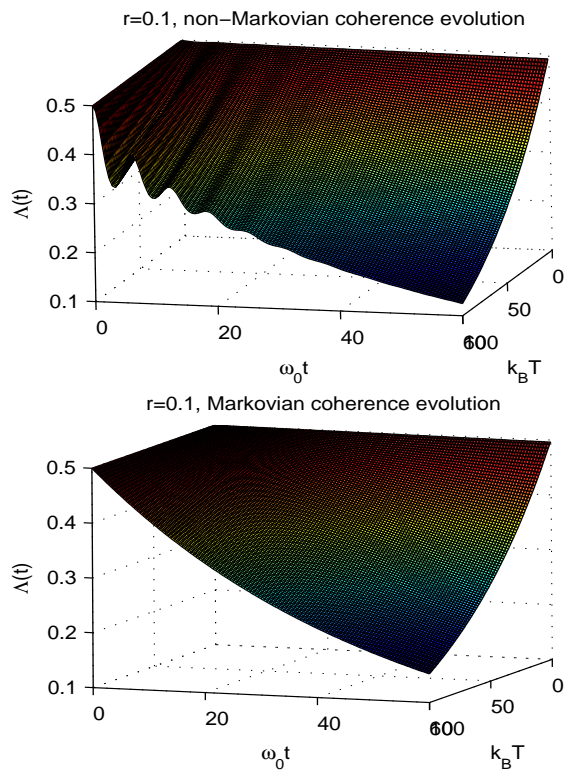}}}
\caption{(Color online) Comparing the non-Markovian coherence
dynamics with the Markovian one by the time evolution of
$\Lambda(t)$ as a function of $k_BT$  for initial state
$(x_0,~y_0,~z_0)=\left(\frac{\sqrt2}{4},~\frac{\sqrt2}{4},~\frac{\sqrt3}{2}\right)$,
$r=0.1$, $\omega_0=1$, and $\alpha^2=0.01$. }
\end{figure}

As we indicated above, temperature is the key factor in the
non-Markovian system coefficients. In Fig. 1, we plot the coherence
function $\Lambda(t)$ vs ``temperature $k_BT$" vs $\omega_0t$ in
$r=0.1$, and the initial state
$(x_0,~y_0,~z_0)=\left(\frac{\sqrt2}{4},~\frac{\sqrt2}{4},~\frac{\sqrt3}{2}\right)$
for the quantum system without control effect,

\begin{equation}\label{master equation3}
\frac{d}{dt}\rho(t)=-i[\hat{H}_s(t), \rho]+\Gamma_1(t)\mathcal
{D}[\sigma^-]\rho+\Gamma_2(t)\mathcal {D}[\sigma^+]\rho.
\end{equation}

 From Fig. 1 we can compare
the non-Markovian coherence dynamics with the Markovian one clearly.
The up figure is the non-Markovian one from which we can see the
oscillation of the $\Lambda(t)$. Moreover, at the low temperature
especially $0$ temperature the non-Markovian effect is faint, as the
temperature rises, the non-Markovian becomes more and more obvious,
while the Markovian one decays exponentially. This phenomenon
embodies the non-Markovian effect, which is evidently different from
the Markovian property. The reason is that due to the non-Markovian
memory effect, particularly $\Delta(t)<0$ in Eq (\ref{Delta}), the
coherence oscillates. With $\Delta(t)-\gamma(t)>0$ the quantum
system coherence descended whilst $\Delta(t)-\gamma(t)<0$ the
coherence ascended. In high temperature the Markovian quantum system
decays exponentially and vanish only asymptotically, but in the
non-Markovian system the coherence $\Lambda(t)$ oscillates, which is
evidently different from the Markovian. In this case the
non-Markovian property becomes evidently. From (\ref{Delta}), we
have learned that in the strong coupling, high temperature and high
cutoff frequency regimes, the considerable back-action of the non-
Markovian reservoir effectively counteracts the dissipation.

\section{Example: quantum decoherence control}

As is introduced in the introduction, quantum computing and quantum
communication have attracted a lot of attention due to their
promising applications such as the speedup of classical computations
and secure key distributions. Although the physical implementation
of basic quantum information processors has been reported recently,
the realization of powerful and useable devices is still a
challenging and yet unresolved task. A major difficulty arises from
the coupling of a quantum system to its environment that leads to
decoherence. Various methods have been proposed to reduce this
unexpected effect in the past decade, which can be divided into
coherent  and incoherent  control, according to how the controls
enter the dynamics. However, the above control strategies render the
quantum systems are always  neglect the quantum measurement
back-action or study simple systems with the Markovian
approximation.
 In this paper, we restrict our discussion to the
no-Markovian open quantum system, and consider its dynamics with
feedback control and therefore the dynamics  obeys the no-Markovian
master equation (\ref{SME3}).

 \begin{figure}
 \centerline{\scalebox{0.6}[0.8]{\includegraphics{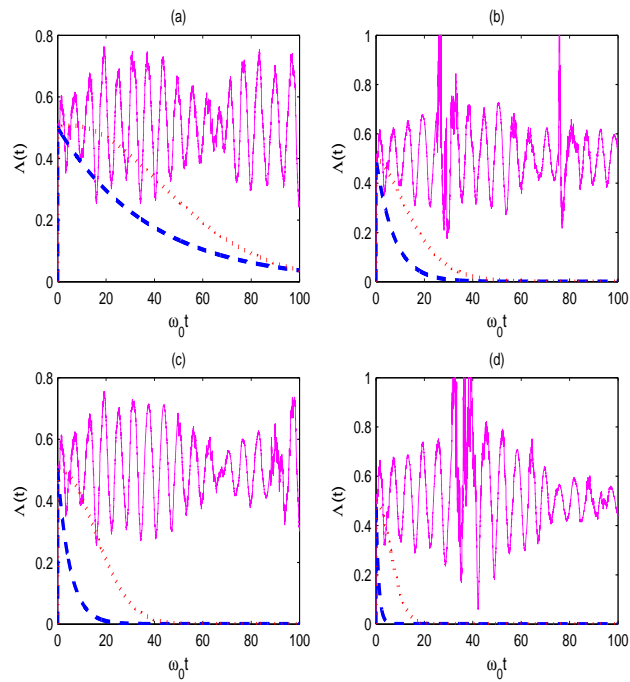}}}
\caption{(Color online) Dynamics of coherence function $\Lambda(t)$
of non-Markovian optimal decoherence control (magenta solid line),
non-Markovian without control (red dotted line), and  Markovian
without control (blue dashed line) for difference cases: (a)
$r=0.5,~~k_BT=1$, (b)$r=3,~~k_BT=1$, (c)$r=0.5,~~k_BT=10$, and (d)
$r=3,~~k_BT=10$ respectively. The other parameters are chosen as
$M=0.05$, and $\eta=1$. }
\end{figure}

To quantify the decoherence dynamics of the qubit, we apply the
following quantity $\Lambda(t)$ which is determined by the
off-diagonal elements of the reduced density matrix.
\begin{equation}
\left|\rho_{12}(t)\right|=\Lambda(t)\left|\rho_{12}(0)\right|.
 \end{equation}
 In the Bloch
 representation, we have
\begin{equation}
\Lambda(t)=\frac{\sqrt{x^2(t)+y^2(t)}}{\sqrt{x^2_0+y^2_0}}.
 \end{equation}
The decoherence factor maintains unity when the reservoir is absent
and vanishes for the case of completely decoherence. For
definiteness, we consider the following initial pure state of the
qubit $ |\psi(0)\rangle=\alpha|+\rangle+\beta|-\rangle.$  To use the
optimal control method (\ref{o1},\ref{o2},\ref{o3},\ref{o4}) we set
the optimal control target $\rho_T$ as the free evolution
$\dot{\rho_T}=-\frac{i}{2}[\omega_0\sigma_z,
 \rho_T].$
It is easy to solve the equation with Bloch representation
$x_T(t)=x_0\cos\omega_0 t-y_0\sin\omega_0 t, y_T(t)=x_0\sin\omega_0
t+y_0\cos\omega_0 t, z_T(t)=z_0. $

To demonstrate the effectiveness of our optimal control strategy, we
present numerical simulations with the initial state
$(x_0,~y_0,~z_0)=\left(\frac{\sqrt2}{4},~\frac{\sqrt2}{4},~\frac{\sqrt3}{2}\right)$,
and coupling constant $\alpha^2=0.01$, optimal control weighting
factor $\theta=1$, and $\omega_0=1$ as the norm unit. Moreover, we
regard the temperature as a key factor in decoherence process.
Another reservoir parameter playing a key role in the dynamics of
the system is the ratio $r=\omega_c/\omega_0$ between the reservoir
cutoff frequency $\omega_c$ and the system oscillator frequency
$\omega_0$. In Fig. 2, we plot  the dynamics of coherence function
$\Lambda(t)$ of non-Markovian optimal decoherence control (magenta
solid line), non-Markovian without control (red dotted line), and
Markovian without control (blue dashed line) for difference cases:
(a) $r=0.5,~~k_BT=1$, (b)$r=3,~~k_BT=1$, (c)$r=0.5,~~k_BT=10$, and
(d) $r=3,~~k_BT=10$ respectively. The other parameters are chosen as
$M=0.05$, and $\eta=1$. From Fig. 2, it is worth noting that as
increasing the ratio $r$, or increasing the temperature $k_BT$ the
coherence lasting time becomes shorter and shorter. From Fig.1 we
have known that non-Markovian reservoir has dual effects on the
qubit: dissipation and back-action. The dissipation effect exhausts
the coherence of the qubit, whereas the back-action one revives it.
Here we still can observe the dynamical mechanism of the
non-Markovian effect: the non-Markovian system's coherence lasting
time is always longer than the Markovian one. From the simulation
results, the coherence function $\Lambda(t)$ will be completely lost
in the absence of control neither non-Markovian system (dotted line)
nor Markovian system (dashed line). However, the feedback control
steers it to a stationary stochastic process which fluctuates around
the target.

\section{Conclusions}

In conclusion, we have investigated the problem of optimal control
of non-Markovian open quantum system via feedback. At first we
analyzed the non-Markovian quantum system and the master equation.
In general, the reduction of the degrees of freedom in the effective
description of the open system results in non-Markovian behavior.
The non-Markovian master equation for the quantum system thus
supports to investigation of non-Markovian effects beyond the
Born-Markovian approximation. In this paper we make a thorough
examination of the difference between the Markovian system and the
non-Markovian one. The main difference is that in the non-Markovian
master equation, one or several of the dissipation coefficients
become temporarily negative which expresses the presence of strong
memory effects in the reduced system dynamics. From our analytic and
numerical results, we find that the non-Markovian reservoir has dual
effects on the qubit: dissipation and backaction. The dissipation
effect exhausts the coherence of the qubit, whereas the backaction
one revives it. In the strong coupling, high temperature and high
cutoff frequency regimes, the considerable backaction of the
non-Markovian reservoir effectively counteracts the dissipation.

Based on the non-Markovian master equation we analyzed the optimal
control problem via feedback. For the quantum system is typically
different to the classical one, the quantum measurement changes the
dynamical evolution. Hence, we consider the quantum weak measurement
and the corresponding master equation is the stochastic one.
Moreover, we designed the control Hamiltonian with the control laws
attained by the stochastic optimal control problem and the
corresponding optimal principle. Usually this kind of problem is
difficult to be analytically solved. We considered this problem in
the non-Markovian two-level system. Through transforming its master
equation into the Bloch vector representation we obtained the
corresponding differential equation with two-sided boundary values.

At last, we considered the exact decoherence dynamics of a qubit in
a dissipative reservoir composed of harmonic oscillators, and
demonstrated the effectiveness of our optimal control strategy.
Obviously, the coherence function will be completely lost in the
absence of control neither  non-Markovian system nor Markovian
system. However, the feedback control steers it to a stationary
stochastic process which fluctuates around the target. In this case
the decoherence can be controlled effectively, which may indicates
that the decoherence rate can be slowed down and decoherence time
can be delayed through design engineered reservoirs.

\section*{Acknowledgments}
This work was supported by the National Natural Science Foundation
of China (No. 60774099, No. 60821091), and the Chinese Academy of
Sciences (KJCX3-SYW-S01).

\end{document}